\documentclass[epj]{mysvjour}
\usepackage{graphicx}
\newcommand{\be}{\begin{equation}}
\newcommand{\ee}{\end{equation}}

\newcommand{\chiPT}{$\chi$PT }
\newcommand{\FRRchiPT}{FRR $\chi$PT }

\begin{document}
\title{
%
%  Comment the following six lines for submission to Archive
%
%\vspace{-2.6cm}
%\hfill \rm \null \hfill
%\hbox{\normalsize ADP-04-???/T???} \\
%\vspace{-2mm}
%\hfill \hbox{\normalsize DESY 04-???} \\
%\vspace{1.65cm}
%
Systematic uncertainties in the precise determination of the strangeness
magnetic moment of the nucleon}
\titlerunning{Systematic uncertainties in the precise determination$\ldots$}
\author{D.~B.~Leinweber\inst{1}\inst{2} \and
        S.~Boinepalli\inst{1} \and
        A.~W.~Thomas\inst{1}\inst{2} \and
        A.~G.~Williams\inst{1} \and
        R.~D.~Young\inst{1}\inst{2} \and
        J.~B.~Zhang\inst{1} \and 
        J.~M.~Zanotti\inst{3}
}
\authorrunning{D.~B.~Leinweber {\it et al.}}
\institute{Special Research Center for the
           Subatomic Structure of Matter, and
           Department of Physics, University of Adelaide,\\
           Adelaide SA 5005,  Australia
\and
           Jefferson Lab, 12000
           Jefferson Ave., Newport News, VA 23606, USA
\and
          John von
          Neumann-Institut f\"ur Computing  NIC,
          Deutsches Elektronen-Synchrotron DESY, D-15738 Zeuthen,
          Germany
}
\date{Received: 3 December 2004}
\abstract{
Systematic uncertainties in the recent precise determination of the
strangeness magnetic moment of the nucleon are identified and
quantified.  In summary, $G_M^s = -0.046 \pm 0.019\ \mu_N$.
}
\maketitle

\section{Introduction}

Recent low-mass lattice QCD simulation results combined with new
chiral extrapolation techniques and the principle of charge symmetry
have enabled a precise determination of the strangeness magnetic
moment of the nucleon $G_M^s$ \cite{Leinweber:2004tc}.  In this paper,
the systematic errors of the approach are explored and quantified.  In
particular, we examine the sensitivity of $G_M^s$ and the magnetic
moments of the baryon octet to the regulator-mass scale of
finite-range regularized chiral effective field theory, the lattice
scale determination, the finite-volume of the lattice, and the
quenched approximation.

\section{Charge Symmetry}

\begin{figure}[tbp]
\vspace*{0.2cm}
\begin{center}
{\includegraphics[height=3.9cm,angle=90]{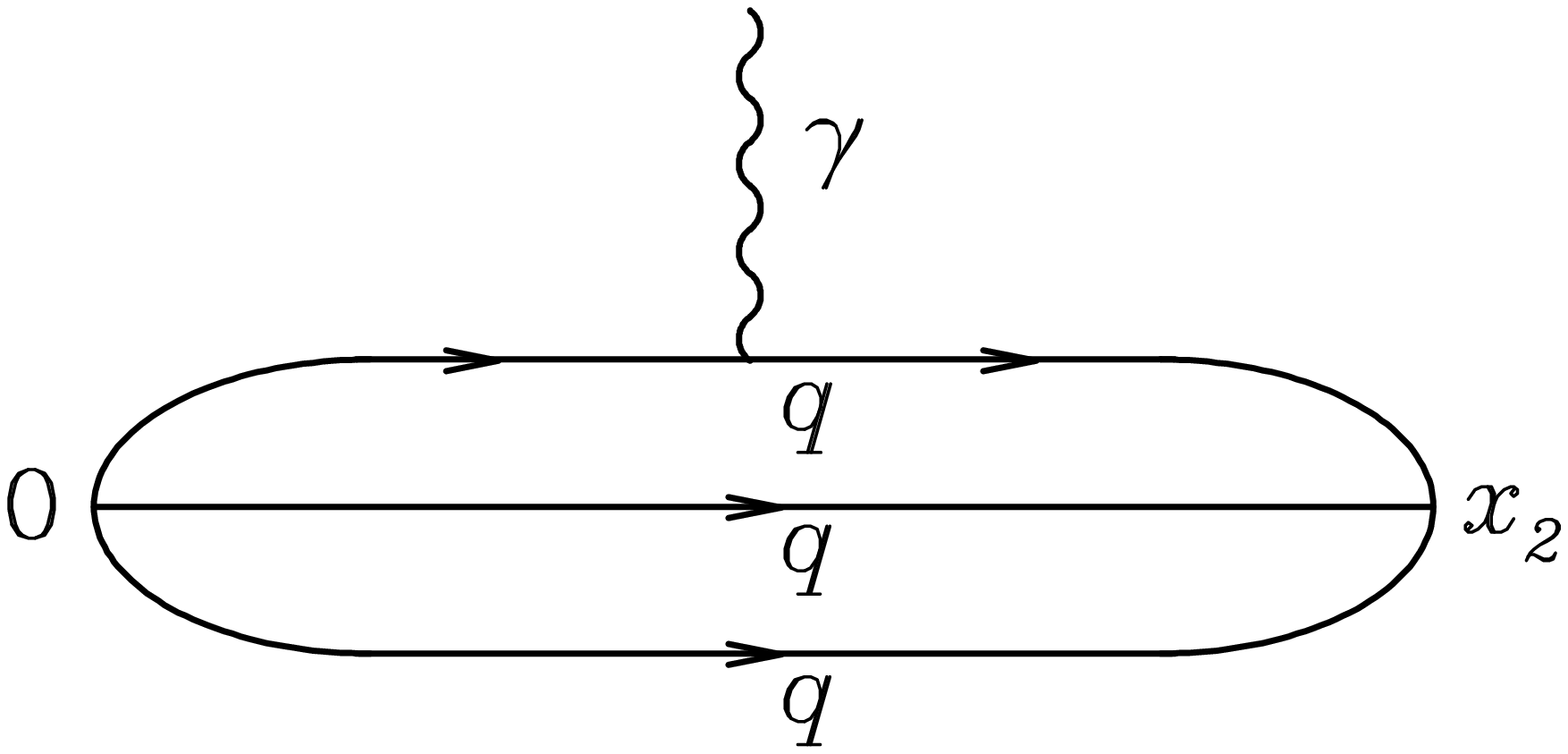} \hspace{0.4cm}
 \includegraphics[height=3.9cm,angle=90]{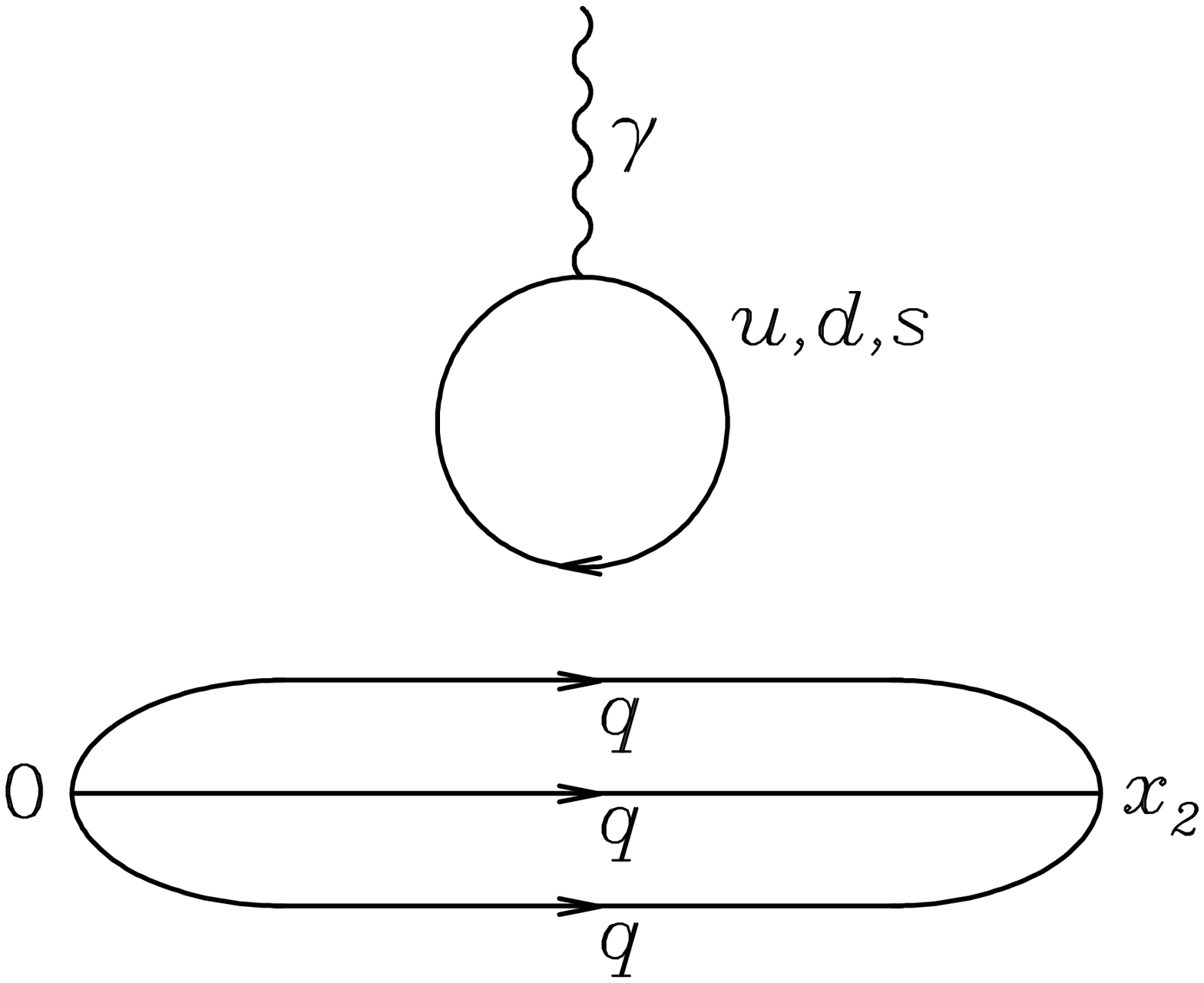}}
\end{center}
%\vspace{-0.7cm}
\caption{Diagrams illustrating the two topologically different
insertions of the current within the framework of lattice QCD.  
These skeleton diagrams for the connected
(left) and disconnected (right) current insertions are dressed by
an arbitrary number of gluons and quark loops.
}
\label{topology}
\end{figure}

The approach \cite{Leinweber:2004tc} centres around two equations for
the strangeness magnetic moment of the nucleon, $G_M^s$, obtained from
charge symmetry
{\small
\begin{equation}
G_M^s = \left ( {\,{}^{\ell}R_d^s \over 1 - \,{}^{\ell}R_d^s }
\right ) \left [ 2 p + n - {u^p \over u^{\Sigma}} \left ( \Sigma^+ -
\Sigma^- \right ) \right ] , 
\label{GMsSigma}
\end{equation}
\begin{equation}
G_M^s = \left ( {\,{}^{\ell}R_d^s \over 
1 - \,{}^{\ell}R_d^s } \right ) \left [
p + 2n - {u^n \over u^{\Xi}} \left ( \Xi^0 - \Xi^- \right ) 
 \right ] .
\label{GMsXi}
\end{equation}
}
%\hspace{-15pt}
Here the baryon labels represent the experimentally measured baryon
magnetic moments and ${}^{\ell}R_d^s \equiv
{G_M^s}/{{}^{\ell}G_M^d}$ is the ratio of $s$- and $d$-quark
sea-quark loop contributions, depicted in the right-hand diagram of
Fig.~\ref{topology}.  ${}^{\ell}R_d^s$ lies in the range (0,1).  The
ratios $u^p / u^{\Sigma}$ and $u^n / u^{\Xi}$ are ratios of
valence-quark contributions to baryon magnetic moments in full QCD as
depicted in the left-hand diagram of Fig.~\ref{topology}.  The latter
are determined by lattice QCD calculations
\cite{Leinweber:2004tc,Zanotti:2001yb,Leinweber:2002bw,FLICscaling,FLIClqm}.

\begin{figure}[tbp]
\begin{center}
{\includegraphics[height=\hsize,angle=90]{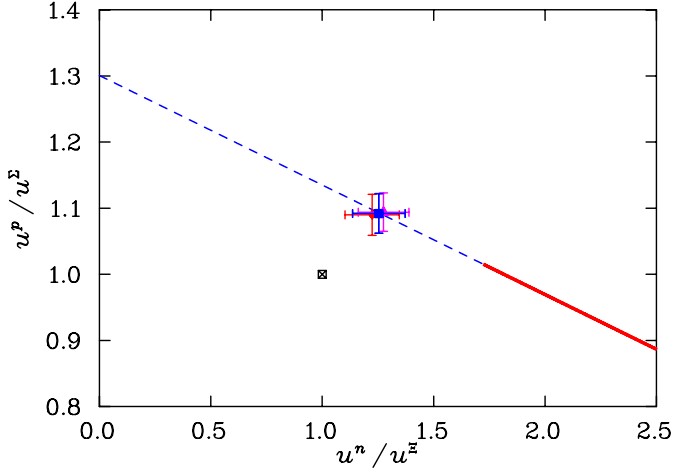}}
\end{center}
%\vspace*{-1.2cm}
\caption{The straight line (dashed $G_M^s(0) < 0$, solid $G_M^s(0) >
0$) indicates the constraint on the ratios $u^p/u^{\Sigma}$ and
$u^n/u^{\Xi}$ implied by charge symmetry and the experimentally
measured magnetic moments.  The assumption of environment independent
quark moments is indicated by the square.  The dependence of the
extrapolated ratios from lattice QCD simulations on the parameter
$\Lambda = 0.7$, 0.8 and 0.9 GeV, (governing the size of
pion-cloud corrections)
% associated with the finite-volume of the
%lattice and artifacts of the quenched approximation 
is illustrated by the cluster of points with $\Lambda$ increasing from
left to right.}
\label{SelfConsFRRchiPTLambda}
\end{figure}

Equating Eqs.~(\ref{GMsSigma}) and (\ref{GMsXi}) provides a linear
relationship between $u^p/u^{\Sigma}$ and $u^n/u^{\Xi}$ which must be
obeyed within QCD under the assumption of charge symmetry -- itself
typically satisfied at the level of 1\% or better \cite{ChargeSymm}.
There are no other systematic uncertainties associated with this
constraint.  Figure \ref{SelfConsFRRchiPTLambda} displays this
relationship by the dashed and solid line.  Since this line does not
pass through the point $(1.0, 1.0)$, corresponding to the simple quark
model assumption of universality, there must be an environment effect
exceeding 12\% in both ratios or approaching 20\% or more in at least
one of the ratios.  To determine the sign of $G_M^s$, it is sufficient
to determine where on this constraint curve, QCD resides.

\section{$u^p/u^{\Sigma}$ and $u^n/u^{\Xi}$ Determination}

Our present precise analysis has been made possible by a significant
breakthrough in the regularization of the chiral-loop contributions to
hadron observables \cite{Young:2002ib,Leinweber:2003dg,Young:2004tb}.
Through the process of regulating loop integrals via a finite-range
regulator (FRR) \cite{Leinweber:2003dg,Donoghue:1998bs}, the chiral
expansion is effectively re-summed to produce an expansion with vastly
improved convergence properties.  

The chiral expansion for the $u$-quark contribution to the proton
magnetic moment in quenched QCD (QQCD), has the form
\begin{eqnarray}
u^p &=& a_0^\Lambda 
            + u^p \, \chi_{\eta'} \, I_{\eta'}(m_\pi, \Lambda)
            + \chi_{\pi B} \, I_B(m_\pi, \Lambda) \nonumber \\
       &&   + \chi_{K B} \, I_B(m_K, \Lambda)
            + a_2^\Lambda \, m_\pi^2 
            + a_4^\Lambda \, m_\pi^4 ,
\label{chiExp}
\end{eqnarray}
where the repeated index, $B$, sums over allowed baryon octet and
decuplet intermediate states.  The dependence of the unrenormalized
coefficients, $a_i^\Lambda$, and the associated dipole-vertex regulated
loop integrals, $I(m_\pi, \Lambda)$, on the regulator parameter,
$\Lambda$, is emphasized by the explicit appearance of $\Lambda$.  The
loop integrals are defined as
%in Ref.~\cite{Leinweber:2004tc}.
%
{\small
\begin{eqnarray}
% &&\hspace*{-5mm}I_\pi(m_\pi,\Lambda) = -\frac{4}{3\pi}\int_0^\infty dk\,
% \frac{k^4}{(k^2+m_\pi^2)^2}u^2(k,\Lambda) \\
&&\hspace*{-5mm}I_B(m, \Lambda)    = 
\\
&&\hspace*{-2mm}-\frac{2}{3\pi} \int  dk 
\frac{ \left (2 \sqrt{k^2+m^2} + \Delta_{BN} \right )\, k^4\, u^2(k,\Lambda)}{
\left (k^2+m^2 \right )^{3/2}\, \left (\sqrt{k^2+m^2} +
\Delta_{BN} \right )^2} \, ,\nonumber\\
&&\hspace*{-5mm}I_{\eta'}(m_\pi, \Lambda) = - \int_0^\infty dk\,
\frac{k^4}{(k^2+m_\pi^2)^\frac{5}{2}}u^2(k,\Lambda) \, ,
\end{eqnarray}
}%
%
%\hspace{-12pt}
where $\Delta_{BN}$ is the relevant baryon mass splitting and the
function $u(k,\Lambda)$ is the dipole-vertex regulator.
The coefficients, $\chi$, denote the known model-independent
coefficients of the non-analytic terms for $\pi$ and $K$ mesons in
QQCD \cite{Leinweber:2002qb,Savage:2001dy}.

\begin{figure}[tp]
\begin{center}
{\includegraphics[height=\hsize,angle=90]{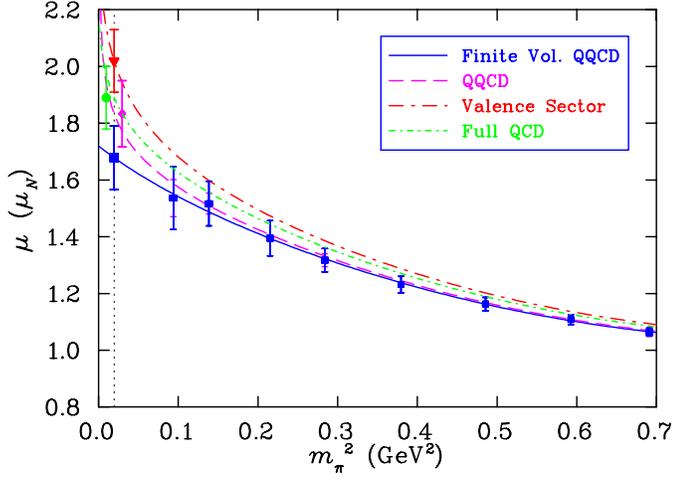}}
\end{center}
%\vspace*{-1.2cm}
\caption{The contribution of a single $u$ quark (with unit charge) to
  the magnetic moment of the proton.  Lattice simulation results
  (square symbols for $m_\pi^2 > 0.05\ {\rm GeV}^2$) are extrapolated to the
  physical point (vertical dashed line) in finite-volume QQCD as well
  as infinite volume QQCD.  Estimates of the valence $u$ quark
  contribution in full QCD and the full $u$-quark sector contribution
  in full QCD are also illustrated.  Extrapolated values at the
  physical pion mass (vertical dashed line), are offset for clarity.
  }
\label{uProton}
\end{figure}

Figure \ref{uProton} illustrates a fit of FRR, quenched chiral
perturbation theory ($\chi$PT) to our lattice results
(solid curve), where only the discrete momenta allowed in the finite
volume of the lattice are summed in evaluating the chiral loop
integrals.  The long-dashed curve that also runs through the lattice
results corresponds to replacing the discrete momentum sum by the
infinite-volume, continuous momentum integral.  For all but the
lightest quark mass, finite volume effects are negligible.

\begin{figure}[tp]
\begin{center}
{\includegraphics[height=\hsize,angle=90]{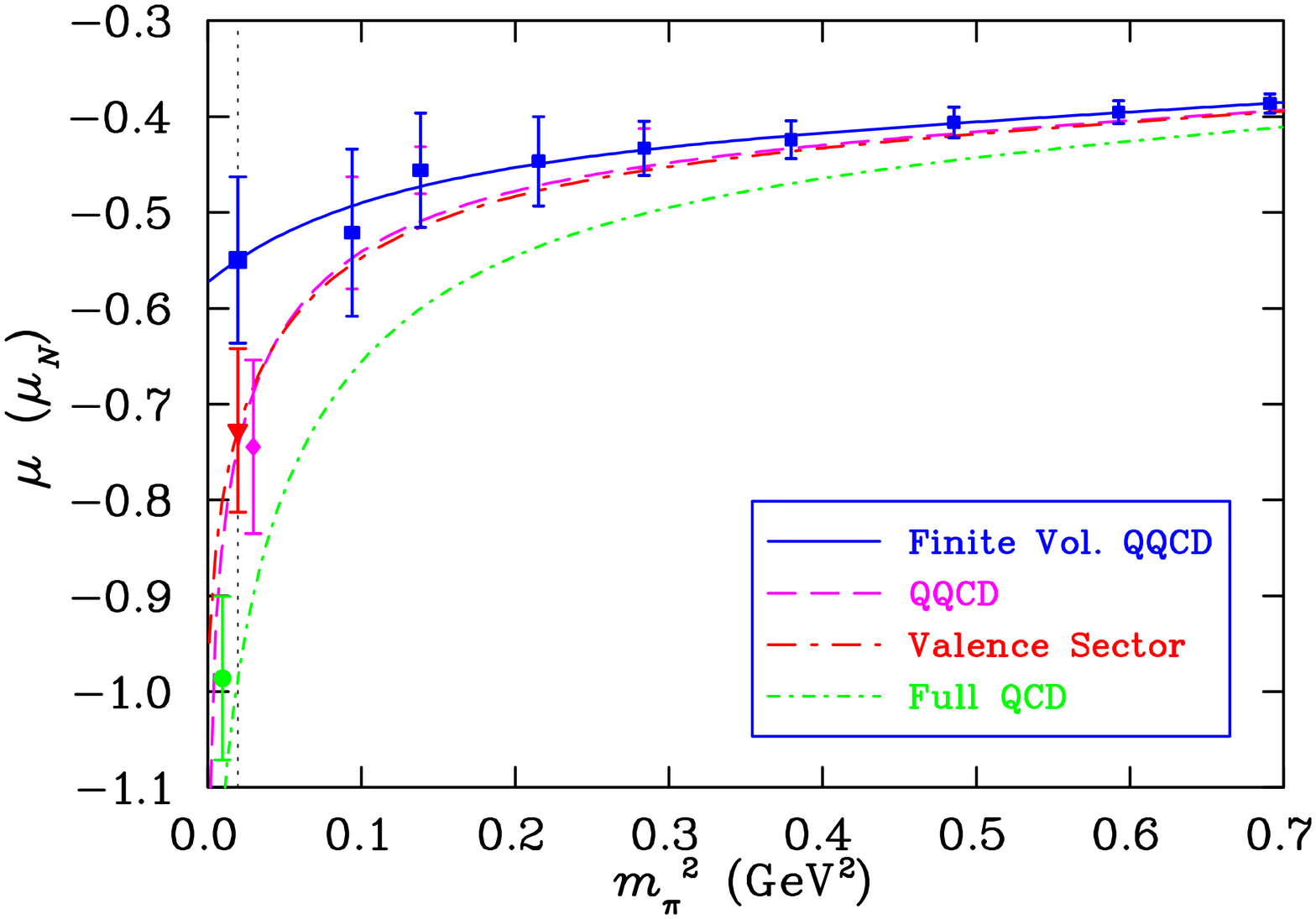}}
\end{center}
%\vspace*{-1.2cm}
\caption{The contribution of the $u$ quark (with unit charge) to the
magnetic moment of the neutron.  Curves and symbols are as described
in Fig.~\protect\ref{uProton}.  }
\label{uNeutron}
\end{figure}

\begin{figure}[tp]
\begin{center}
{\includegraphics[height=\hsize,angle=90]{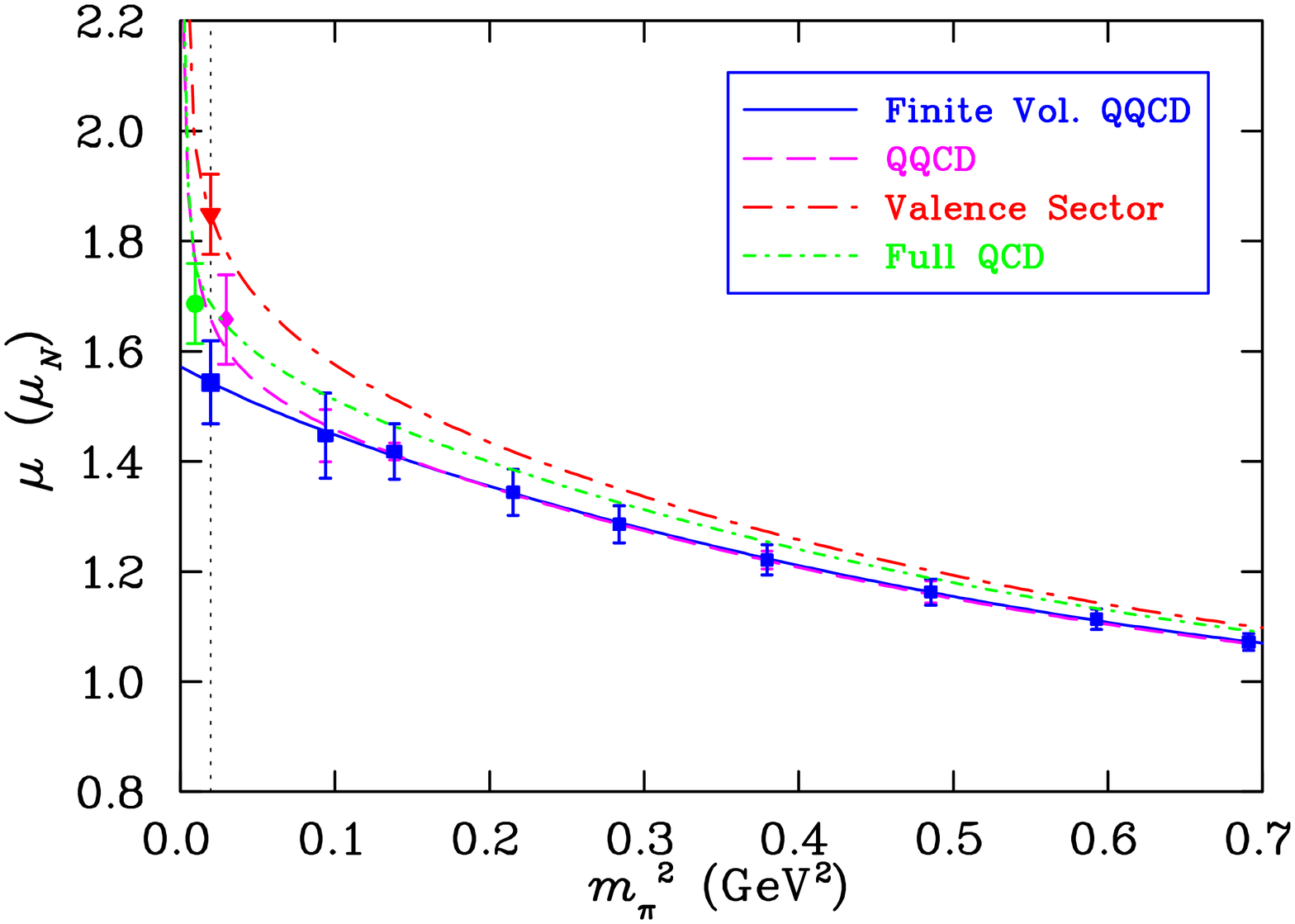}}
\end{center}
%\vspace*{-1.2cm}
\caption{ The contribution of a single $u$ quark (with unit charge) to
the magnetic moment of $\Sigma^+$.  Curves and symbols are as for
Fig.~\protect\ref{uProton}.  }
\label{uSigma}
\vspace{2.4cm}
%end{figure}
%
%begin{figure}[tp]
\begin{center}
{\includegraphics[height=\hsize,angle=90]{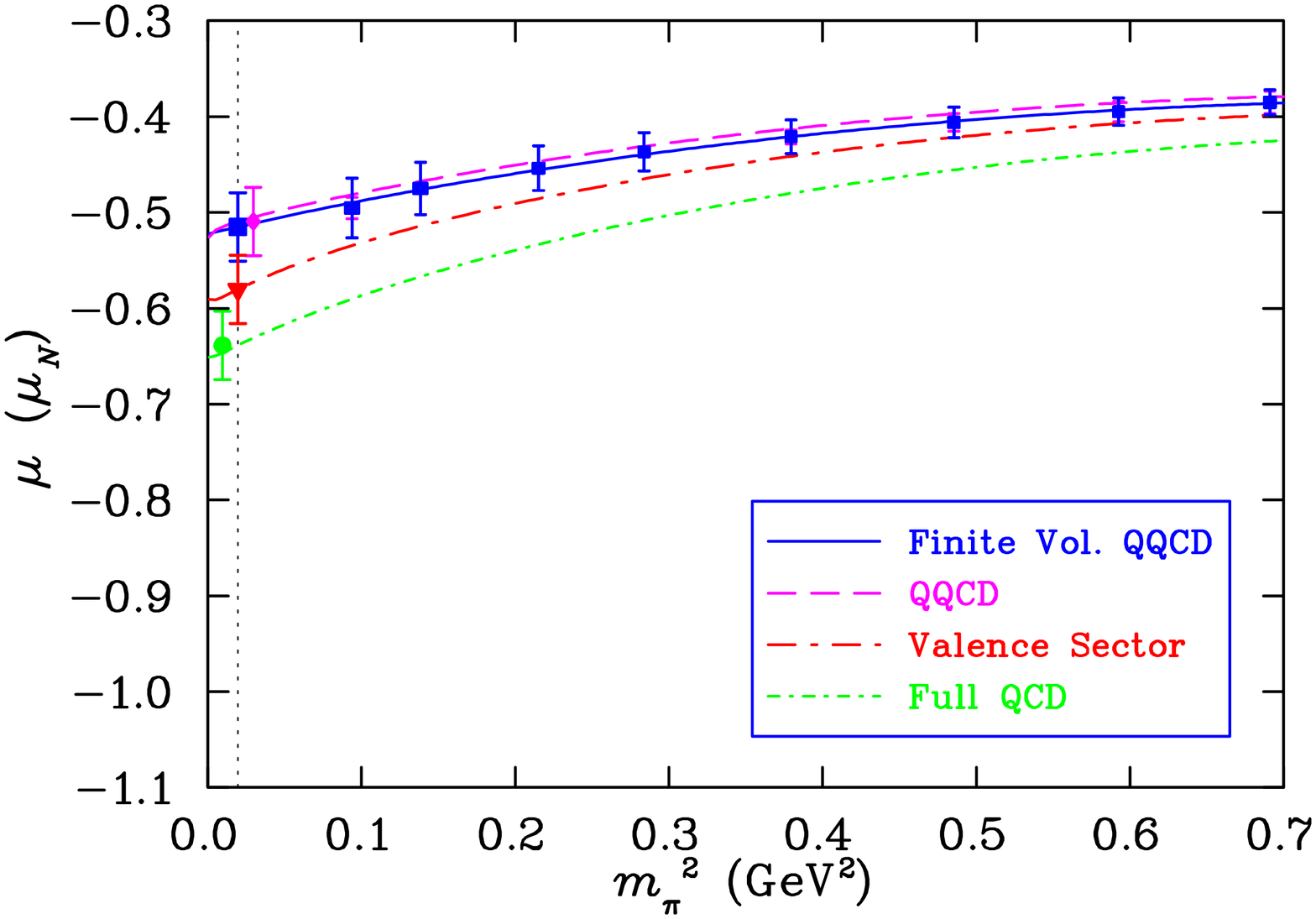}}
\end{center}
%\vspace*{-1.2cm}
\caption{ The contribution of the $u$ quark (with unit charge) to the
magnetic moment of the $\Xi^0$ hyperon.  Curves and symbols are as
described in Fig.~\protect\ref{uProton}.  }
\label{uXi}
\end{figure}

Having determined the analytic coefficients $a_{0,2,4}^\Lambda$ for a
particular choice of $\Lambda$, one can correct the chiral properties
of the pion-cloud contribution from QQCD to full QCD
\cite{Young:2004tb,Young:2002cj} by changing the coefficients of the
loop integrals, $\chi_{\eta'}$, $\chi_{\pi B}$, $\chi_{K B}$ of
Eq.~(\ref{chiExp}), to their full QCD counter parts
\cite{Leinweber:2002qb,Savage:2001dy}.  Valence quark contributions in
full QCD are indicated by the long-dash-dot curve in
Fig.~\ref{uProton} (i.e.\ sea-quark loop charges are zero) and the
full $u$-quark sector including the $u$-sea-quark loop contributions
are indicated by the short-dash-dot curve for $\Lambda = 0.8$ GeV.
Figures \ref{uNeutron}, \ref{uSigma} and \ref{uXi} show similar
results for the $u$ quark in $n$, $\Sigma^+$, and $\Xi^0$
respectively.

The importance of correcting for both finite-volume and quenching
artifacts is illustrated in Figs.~\ref{MagMomDiscPos} and
\ref{MagMomDiscNeg}, where the one standard deviation agreement
between the chirally corrected lattice QCD simulation results and the
experimentally measured baryon magnetic moments is highlighted.

\section{Systematic Errors}

\subsection{Regulator Dependence}

\begin{figure}[tbp]
\begin{center}
{\includegraphics[height=\hsize,angle=90]{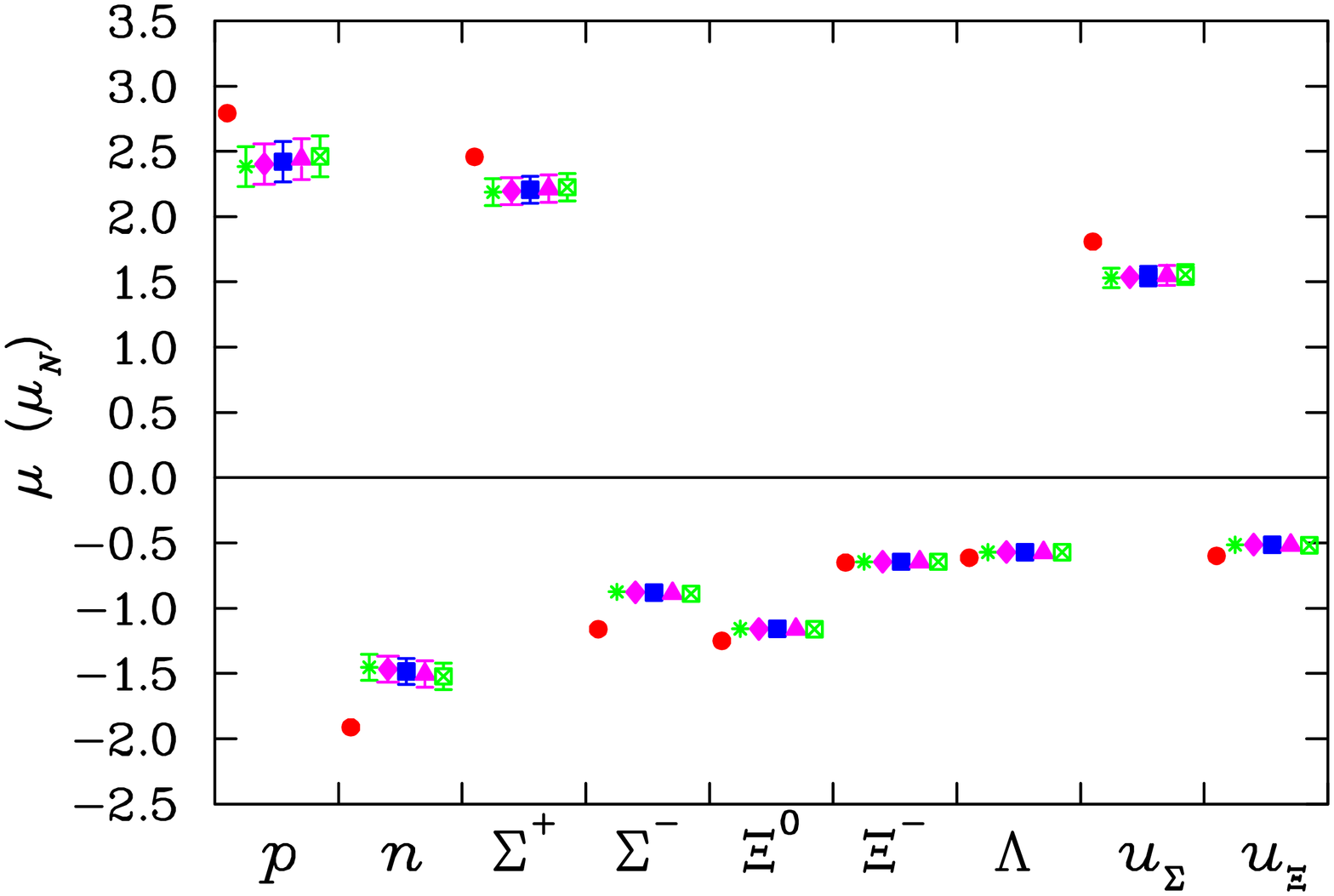}}
\end{center}
%\vspace*{-1.2cm}
\caption{The \FRRchiPT regulator mass dependence of finite-volume
  quenched chiral effective field theory.  Experimental measurements
  are illustrated at left for each baryon for reference.  Results for
  $\Lambda = 0.6$, 0.7, 0.8, 0.9 and 1.0 GeV are illustrated from left
  to right for each baryon.  The small systematic dependence on
  $\Lambda$ relative to the statistical error bars illustrated
  indicate the order of the chiral expansion is adequate for this
  analysis. 
}
\label{MagMomLambdaDisc}
\end{figure}

It is important to investigate systematic errors associated with the
regulator-mass dependence of \FRRchiPT.  The extrapolated results of
finite-volume quenched chiral effective field theory should be
insensitive to the choice of regulator parameter.  When working to
sufficient order in the chiral expansion, changes in the regulation of
loop integrals should be absorbed by changes in the unrenormalized
coefficients, $a_{0,2,4}^\Lambda$, in a manner which preserves the
invariant renormalized coefficients.  The latter are reflected in
Fig.~\ref{MagMomLambdaDisc} which illustrates the insensitivity of
quenched baryon magnetic moments in a finite-volume to the regulator
parameter $\Lambda$.  This systematic error is small relative to the
statistical error.

\begin{figure}[tbp]
\begin{center}
{\includegraphics[height=\hsize,angle=90]{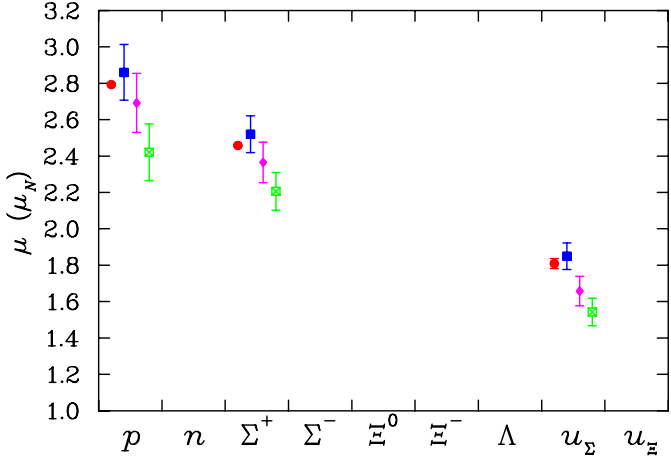}}
\end{center}
%\vspace*{-1.2cm}
\caption{The one standard deviation agreement between the chirally
  corrected lattice QCD simulation results (square symbols) and the
  experimentally measured baryon magnetic moments (circular symbols)
  having positive values.
  Finite-volume quenched results (crossed boxes) and infinite-volume
  quenched results (diamonds) are also illustrated to highlight the
  importance of correcting for both finite-volume and quenching
  artifacts.}
\label{MagMomDiscPos}
%\vspace{1.0cm}
\end{figure}

\begin{figure}[tbp]
\begin{center}
{\includegraphics[height=\hsize,angle=90]{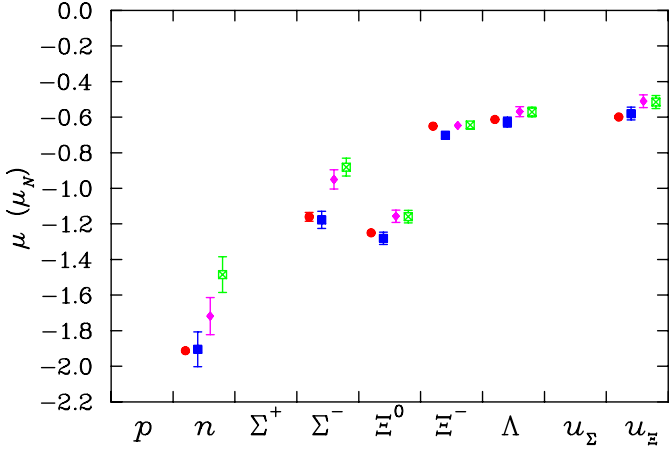}}
\end{center}
%\vspace*{-1.2cm}
\caption{The one standard deviation agreement between the chirally
  corrected lattice QCD simulation results and the experimentally
  measured baryon magnetic moments having negative values.  Symbols
  are as in Fig.~\protect\ref{MagMomDiscPos}.  }
\label{MagMomDiscNeg}
%\vspace{2.0cm}
\end{figure}

\begin{figure}[tbp]
\begin{center}
%{\includegraphics[height=\hsize,angle=90]{MagMomLambda.ps}}
{\includegraphics[height=\hsize,angle=90]{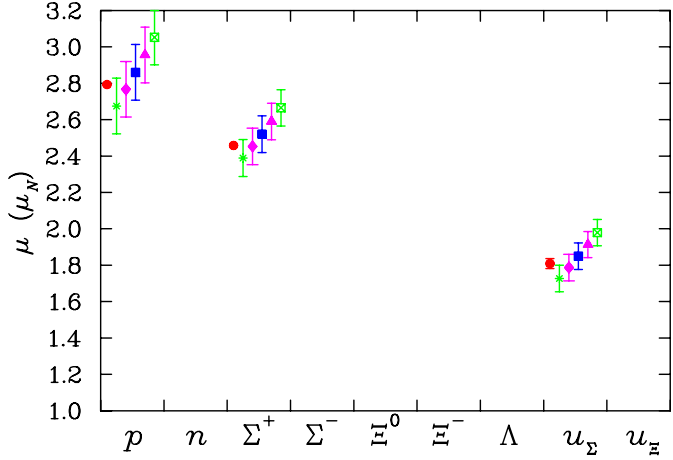}}
\end{center}
%\vspace*{-1.2cm}
\caption{The dependence of positive octet-baryon magnetic moments on the
  parameter $\Lambda = 0.6$, 0.7, 0.8, 0.9 and 1.0 GeV, governing the
  size of pion-cloud corrections associated with the finite-volume of
  the lattice and artifacts of the quenched approximation.
  Experimental measurements, illustrated at left by the filled circle
  for each baryon, indicate that optimal corrections are obtained for
  $0.6 \le \Lambda \le 0.9$ GeV.  }
\label{MagMomLambdaPos}
\end{figure}

\begin{figure}[tbp]
\begin{center}
{\includegraphics[height=\hsize,angle=90]{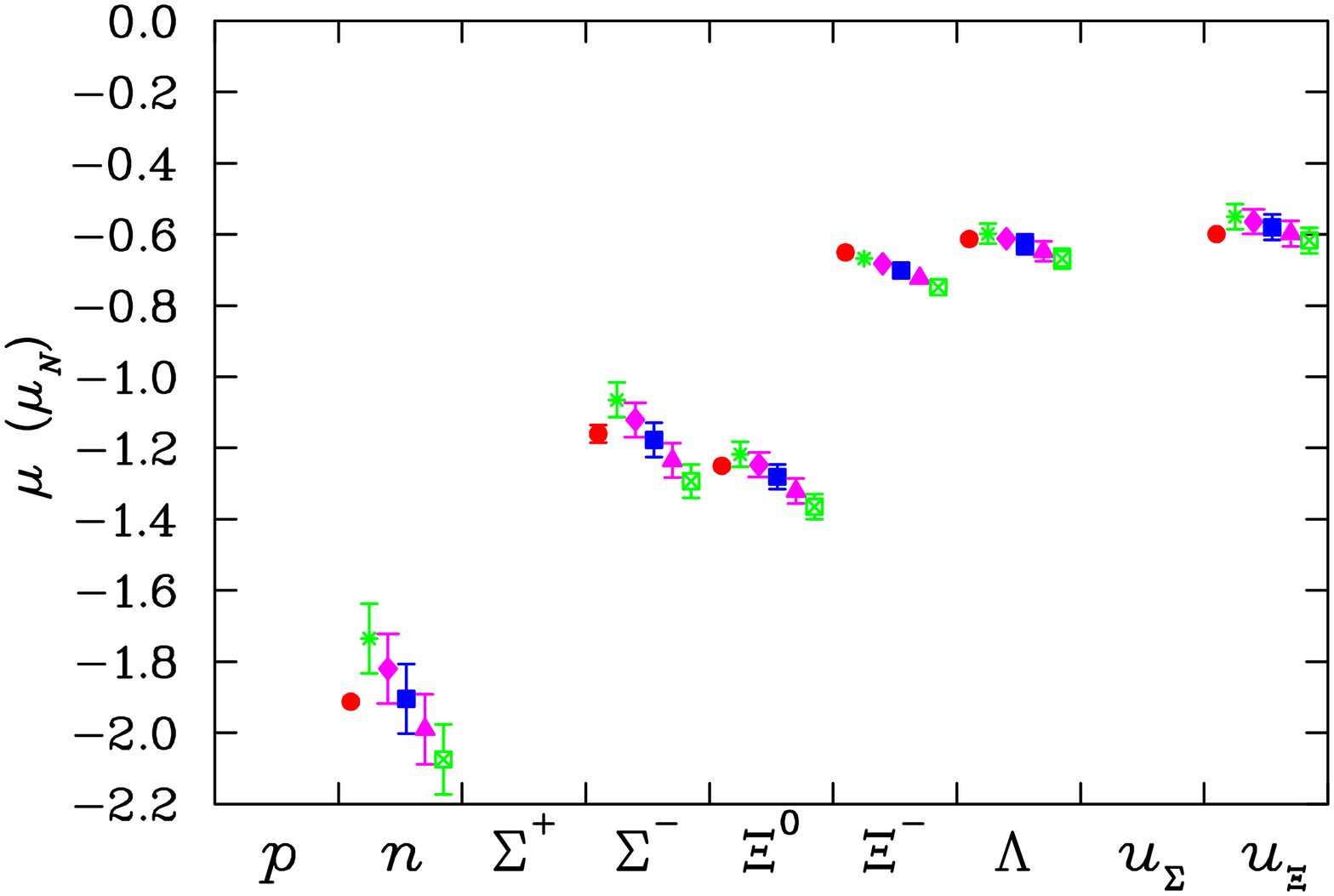}}
\end{center}
%\vspace*{-1.2cm}
\caption{The dependence of negative octet-baryon magnetic moments on the
  parameter $\Lambda = 0.6$, 0.7, 0.8, 0.9 and 1.0 GeV, governing the
  size of pion-cloud corrections associated with the finite-volume of
  the lattice and artifacts of the quenched approximation.
  Experimental measurements, illustrated at left by the filled circle
  for each baryon, indicate that optimal corrections are obtained for
  $0.7 \le \Lambda \le 0.9$ GeV.  }
\label{MagMomLambdaNeg}
\vspace{0.5cm}
\end{figure}

Since the finite-volume and quenching corrections are applied only to
the loop integral contributions, the final results are $\Lambda$
dependent.  In this case, the regulator of the loop integral has
become a model for the axial-vector form factor of the nucleon,
describing the coupling of pions to a core described by the analytic
terms of the FRR expansion.  This approach describes the relation
between quenched and full QCD $N$ and $\Delta$ mass simulation results
as a function of $m_\pi$ very accurately \cite{Young:2002cj}.  Figures
\ref{MagMomLambdaPos} and \ref{MagMomLambdaNeg} display results for
positive and negative baryon magnetic moments in full QCD
respectively.  One-standard deviation agreement is achieved for $0.7
\le \Lambda \le 0.9$.

Because the strangeness magnetic moment of Eqs.~(\ref{GMsSigma}) and
(\ref{GMsXi}) depends only on ratios of magnetic moments, most of this
$\Lambda$ dependence cancels in the final ratios, as illustrated by
the close clustering of points in Fig.~\ref{SelfConsFRRchiPTLambda}.

\subsection{Role of the Decuplet in \chiPT}

\begin{figure}[tbp]
\begin{center}
{\includegraphics[height=\hsize,angle=90]{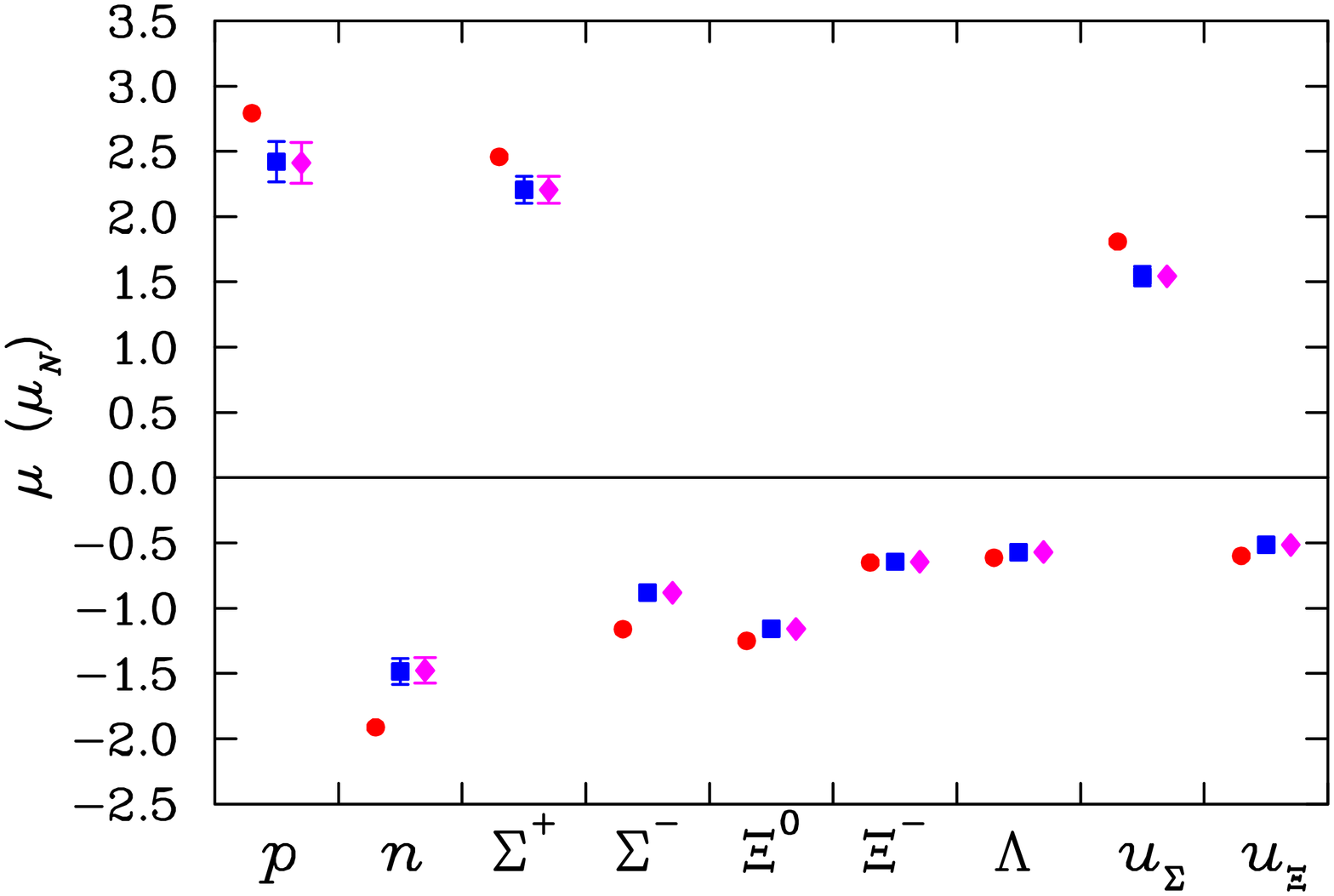}}
\end{center}
%\vspace*{-1.2cm}
\caption{The insensitivity of finite-volume quenched chiral effective
  field theory to NLNA decuplet-baryon contributions.  Results
  including decuplet intermediate states in chiral effective field
  theory (squares) are compared with results excluding the decuplet
  (diamonds).  Experimental measurements (circles), are illustrated at
  left for each baryon for reference.  }
\label{MagMomNoDecDisc}
\vspace{0.55cm}
%\end{figure}
%
%
%\begin{figure}[tbp]
\begin{center}
{\includegraphics[height=\hsize,angle=90]{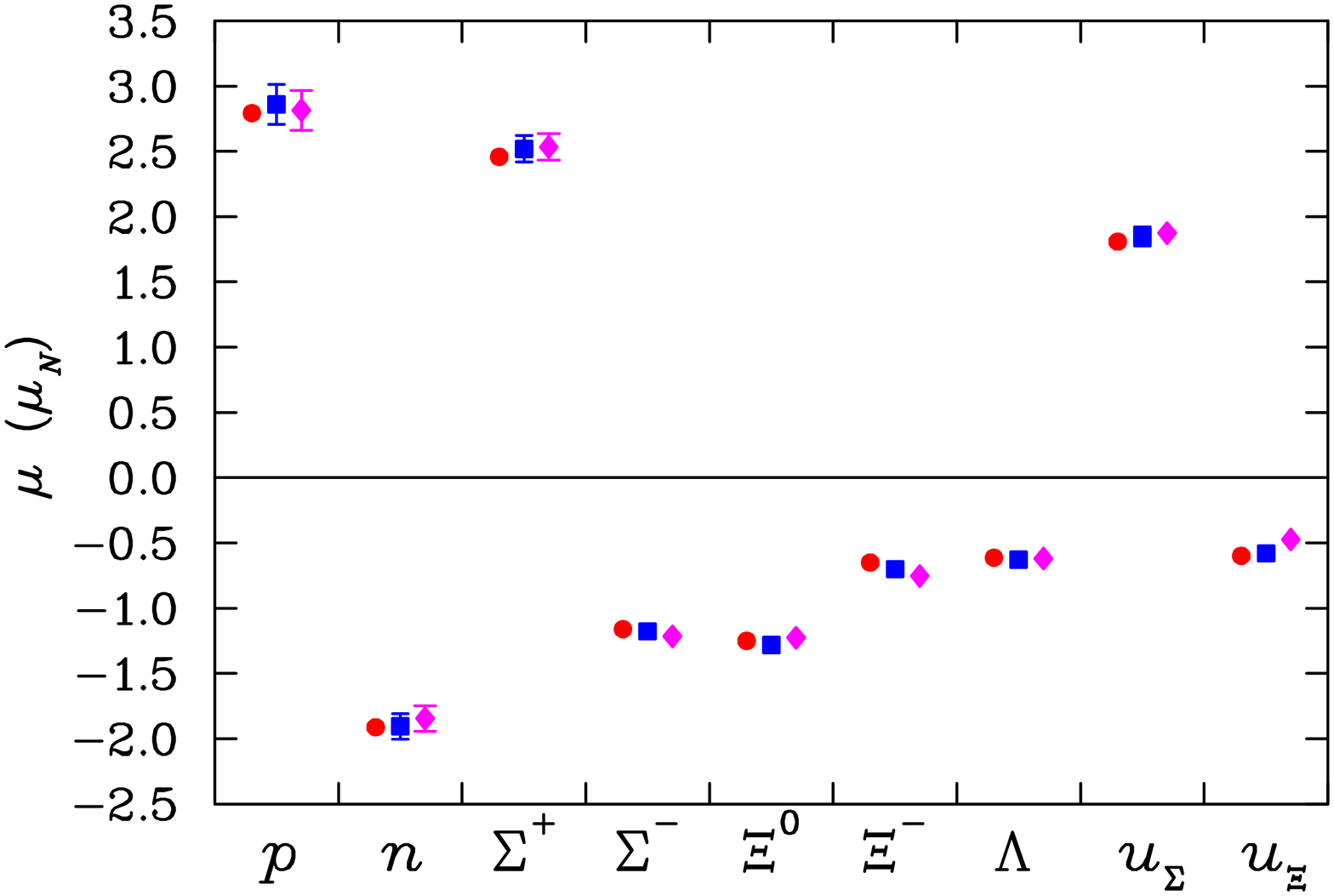}}
\end{center}
%\vspace*{-1.2cm}
\caption{The dependence of octet-baryon magnetic moments on the
  inclusion of NLNA decuplet-baryon contributions in the process of
  correcting the pion cloud of quenched chiral effective field theory.
  Preferred results including decuplet intermediate states in chiral
  effective field theory (squares) are compared with results excluding
  the decuplet (diamonds).  Experimental measurements (circles), are
  illustrated at left for each baryon for reference.  }
\label{MagMomNoDec}
\vspace{0.6cm}
\end{figure}

It is often argued that next-to-leading-order non-analytic (NLNA)
contributions from the baryon decuplet are essential in describing the
mass dependence of nucleon magnetic moments.  While the decuplet
baryon contributions are not necessarily small, we find the
non-analytic curvature induced by these contributions is sufficiently
subtle that it may be accurately absorbed by the analytic terms of the
chiral expansion.  Figure \ref{MagMomNoDecDisc} illustrates the
insensitivity of finite-volume quenched chiral effective field theory
to NLNA decuplet-baryon contributions.

Figure~\ref{MagMomNoDec} confirms that the NLNA decuplet-baryon
contributions are indeed large in some cases and as such are important
in correcting the artifacts of the quenched approximation.  However,
other more highly-excited baryon resonances have small couplings to
the ground-state baryon octet relative to that for the decuplet and
provide negligible corrections.

\subsection{Scale Dependence}

Setting the scale in quenched QCD simulations is somewhat problematic.
Different observables lead to different lattice spacings, $a$.  If one
is explicitly correcting the one-loop pion-cloud contributions to
hadronic observables, as we are here, then clearly one must set the
scale using an observable insensitive to chiral physics.  This
excludes observables such as the rho-meson mass, nucleon mass, or the
pion decay constant commonly used in the literature to hide the
artifacts of the quenched approximation.  On the other hand, the
heavy-quark phenomenology of the static-quark potential provides an
optimal case.  In particular, the Sommer parameter, $r_0$, is ideal as
it sets the scale by equating the force between two static quarks in
QQCD and full QCD at a precise separation of $r_0 = 0.49$ fm.  

\begin{figure}[tbp]
\begin{center}
{\includegraphics[height=\hsize,angle=90]{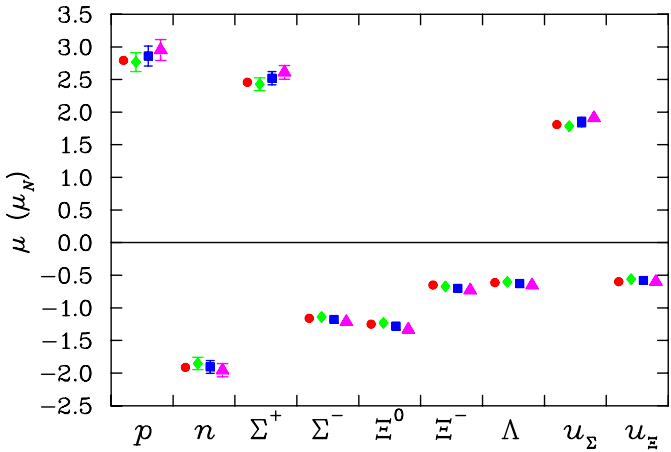}}
\end{center}
%\vspace*{-1.2cm}
\caption{The dependence of octet-baryon magnetic moments on the
  matching criteria for determining the lattice spacing, $a$.  While
  $a_{r_0} = 0.128$ fm set by $r_0$ (square symbols) is the preferred
  method for determining the scale, results for $a_\sigma = 0.134$ fm
  set by the string tension (triangles) and a third spacing of 0.122
  fm (diamonds) are also illustrated.  Experimental measurements
  (filled circles), are illustrated at left for each baryon.  }
\label{MagMomScale}
\end{figure}

\begin{figure}[tbp]
\begin{center}
{\includegraphics[height=\hsize,angle=90]{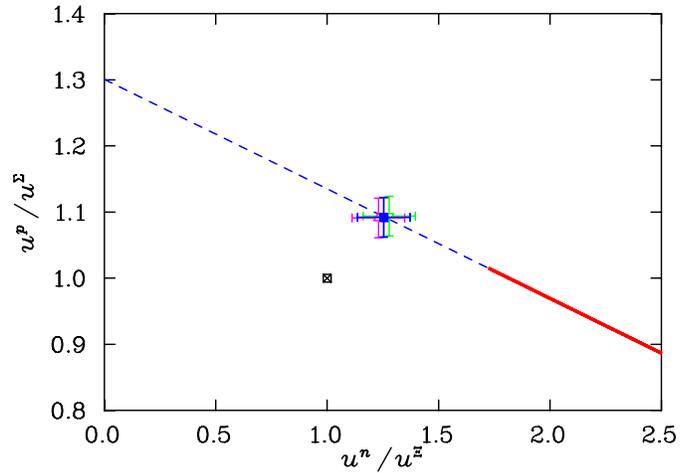}}
\end{center}
%\vspace*{-1.2cm}
\caption{The charge symmetry constraint line (dashed $G_M^s(0) < 0$,
solid $G_M^s(0) > 0$) on the ratios $u^p/u^{\Sigma}$ and
$u^n/u^{\Xi}$.  The dependence of the ratios from chirally-corrected
quenched lattice QCD on the scale parameter, $a = 0.134$, 0.128, and
0.122 fm, is illustrated by the cluster of points with $a$ decreasing
from left to right.}
\label{SelfConsFRRchiPTScale}
\end{figure}

The string tension might also be used, but screening of the potential
in full QCD makes this measure poorly defined.  However, we consider
it here as a measure of the systematic error encountered in setting
the scale of the lattice QCD results.  Figure \ref{MagMomScale}
illustrates this scale dependence on baryon magnetic moments and
Fig.~\ref{SelfConsFRRchiPTScale} illustrates the rather minor impact
this systematic uncertainty has on the valence-quark moment ratios
vital to determining the sign of $G_M^s$.

Collecting the variation of the valence-quark moment ratios from
variations in $\Lambda$ (which maintain one-standard-deviation
agreement with experiment), variations in setting the lattice
scale and statistical errors determined by a third-order
single-elimination jackknife analysis, one finds
\begin{equation}
\frac{u^p}{u^{\Sigma}} = 1.092 \pm 0.030\ {\rm and}\ 
\frac{u^n}{u^{\Xi}} = 1.254 \pm 0.124\, .
\label{finalRatios}
\end{equation}

\begin{figure}[tbp]
\begin{center}
{\includegraphics[height=\hsize,angle=90]{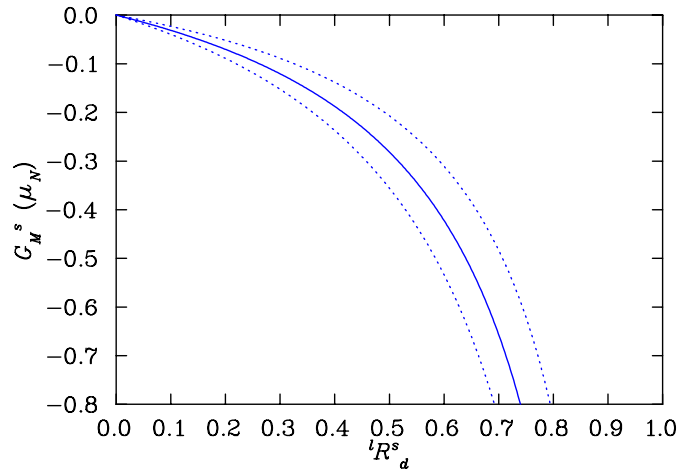}}
\end{center}
%\vspace*{-1.2cm}
\caption{The dependence of $G_M^s$ on the strange to light sea-quark
  loop ratio ${}^{\ell}R_d^s$.  Standard error limits have
  there origin in the systematic error summary of
  Eq.~\protect\ref{finalRatios}. 
}
\label{GMsRsdXiNzoom}
\end{figure}

Using the experimental magnetic moments, one observes that
Eq.~(\ref{GMsXi}) 
{\small
\begin{equation}
G_M^s = \left ( {\,{}^{\ell}R_d^s \over 1 - \,{}^{\ell}R_d^s } \right ) \left [
-1.033 - {u^n \over u^{\Xi}} \left ( -0.599 \right ) \right ] \, ,
\label{great}
\end{equation}
} 
%\hspace{-12pt}
is least sensitive to variation in the valence-quark moment ratio,
and hence provides the most precise determination for $G_M^s$.
Figure \ref{GMsRsdXiNzoom} plots $G_M^s$ as a function of
${}^{\ell}R_d^s$ with standard error limits provided by
Eq.~(\ref{finalRatios}).

\section{Estimating ${}^{\ell}R_d^s \equiv
  {G_M^s}/{{}^{\ell}G_M^d}$}

The symmetry of the three-point correlation function \cite{Leinweber:1996ie}
describing the sea-quark loop contributions to the nucleon, depicted
in the right-hand side of Fig.~\ref{topology}, ensure that the chiral
expansion for this quantity is identical for all three quark flavours,
up to simple charge factors.  For the $d$- or $s$-quark loop
contributions, the only difference that can arise is whether one
evaluates the chiral expansion at the pion or kaon mass.  The leading
non-analytic contribution to the chiral expansion involves two
pseudoscalar meson propagators, and therefore one expects
contributions to ${}^{\ell}R_d^s$ in the ratio $m_\pi^4 / m_K^4 \sim
0.1$.

To be more precise, one can use the same successful (single-parameter)
model, previously used to correct the quenched simulation results to
full QCD, as highlighted in Figs.~\ref{MagMomDiscPos} and
\ref{MagMomDiscNeg}, to provide an estimate for ${}^{\ell}R_d^s$.
Evaluating the loop integrals with $\Lambda = 0.8 \pm 0.2$ GeV
yields ${}^{\ell}R_d^s = 0.139$ with $0.096 \le {}^{\ell}R_d^s \le
0.181$.  This uncertainty dominates the final uncertainty in $G_M^s$.

\begin{table*}[t]
\caption{Sources of uncertainty and their contribution to the
strangeness magnetic moment of the nucleon, $G_M^s$, in units of
nuclear magnetons, $\mu_N$.  Uncertainties are documented for $G_M^s$
obtained from the valence-quark ratio $u^p/u^\Sigma$ in
Eq.~(\protect\ref{GMsSigma}), from the valence-quark ratio $u^n/u^\Xi$
in Eq.~(\protect\ref{GMsXi}) and from a statistically weighted (SW)
average of these two determinations.}
\label{errorBudget}
\newcommand{\m}{\hphantom{$-$}}
\renewcommand{\arraystretch}{1.2} % enlarge line spacing
%\begin{center}
\begin{tabular}{llccc}
\noalign{\smallskip}\hline\noalign{\smallskip}
                    &
&$u^p/u^\Sigma$, Eq.~(\protect\ref{GMsSigma})
&$u^n/u^\Xi$, Eq.~(\protect\ref{GMsXi})
&SW Average\\
Uncertainty Source  &Parameter Range
&$G_M^s = -0.045$ 
&$G_M^s = -0.046$ 
&$G_M^s = -0.046$ \\
\noalign{\smallskip}\hline\noalign{\smallskip}
Statistical Errors &                               &0.016 &0.009 &0.008 \\
Chiral corrections &$0.7 \le \Lambda \le 0.9$ GeV  &0.001 &0.002 &0.002 \\
Scale Determination &$0.122 \le a \le 0.134$ fm    &0.001 &0.002 &0.002 \\
${}^{\ell}R_d^s$ Determination &$0.096 \le {}^{\ell}R_d^s \le 0.181$
                                                   &0.016 &0.017 &0.017 \\
\noalign{\smallskip}\hline\noalign{\smallskip}
Total Uncertainty  &                               &0.023 &0.019 &0.019 \\
\noalign{\smallskip}\hline\noalign{\smallskip}
\end{tabular}\\[2pt]
%\end{center}
\end{table*}

Table \ref{errorBudget} summarizes the sources of uncertainty and
their contributions to the final determination
\begin{equation} 
{G_M^s} = -0.046 \pm 0.019\ \mu_N \, ,
\label{GMs}
\end{equation} 
for the strange quark contribution to the magnetic moment of the
nucleon.

%\newpage
\section*{Acknowledgments}
We thank the Australian Partnership for Advanced Computing (APAC) for
generous grants of supercomputer time which have enabled this project.
Support from the South Australian Partnership for Advanced Computing
(SAPAC) and the National Facility for Lattice Gauge Theory is also
gratefully acknowledged.  DBL thanks Jefferson Lab for their kind
hospitality where the majority of this research was performed.  This
work is supported by the Australian Research Council and by DOE
contract DE-AC05-84ER40150 under which SURA operates Jefferson Lab.

\end{document}